# ICAConfPubs: A Dataset and User Interface for ICA Conference Papers (2003–2018)

Hongtao Hao[1], Xinyue Chen[2], Jiye Sun[3], Yanling Zhao[4], Jing Zhang[5]

[1]Department of Computer Sciences, The University of Wisconsin-Madison, USA
[2]School of Intelligence Science Technology, Peking University, China
[3]The Media School, Indiana University Bloomington, USA
[4]School of Communication, Northwestern University, USA
[5]Department of Communication and Media, the University of Michigan, USA

**Abstract**

This paper presents a comprehensive dataset of past ICA (The International Communication Association) annual conference papers from 2003 to 2018, encompassing 27,466 papers, 21,038 authors, and 4,935 sessions. We made the dataset publicly available in both CSV and JSON formats. Additionally, we developed an API to facilitate programmatic access, and an intuitive user interface to enable users to navigate and explore the data more easily. The web application, API documentation, downloadable data, and reproducible code to obtain and process the data are available at https://ica.hongtaoh.com.

## Contents

ICAConfPubs 2003–2018

Papers Authors Sessions About API

2003-2018 · 27,466 papers · 21,038 authors · 4,935 sessions

Search ICA conference papers.

Enter a keyword, sentence, or paragraph. The system finds the most relevant papers from 27,466 ICA conference papers (2003–2018) using semantic similarity.

I am interested in how social media misinformation circulates during elections, how people correct false claims, and whether partisan identity shapes what audien

top 10

Search papers

I am interested in how social media misinformation circulates during elections, how people correct false claim...

health communication and patient participation

artificial intelligence and communication work

Figure 1: The interface of ICAConfPubs live website and paper details. The headers include papers, authors, and sessions. The authors page contains all authors and click on each author will show all papers authored by that person. The sessions page contains all sessions and clicking on each session will show all papers in that session. Clicking on each paper will navigate to a separate page for a paper where you can see information about authors, conference year, division, session, and abstract. Users can also do semantic search based on their queries.

## Introduction

The year 2026 marks the 76th Annual Conference of the International Communication Association (ICA). Each year, the annual conference welcomes submissions from scholars around the world. As the largest and most prominent association in the field of Communication, the ICA annual conference serves as a venue for worldwide scholars to present their most recent research progress and exchange ideas.

Unfortunately, all these data, i.e., submitted papers, participating scholars, and organized sessions, are not readily available to the public. Annual conferences from 2003 to 2018 have official websites that are powered by the conventions system by `allacademic.com`. Conference programs from 2019 and onward are only available in PDF formats to the public. Conferences programs prior to 2003 were not available. Even though online programs exist, they are complicatedly structured and not ready to organize and analyze for the public and scholars alike.

ICA annual conferences data are valuable and useful because they can:

- Inspire new research ideas. Right now, most communication literature comes from journal papers (searched mostly in Google Scholar). Findings from conferences may provide a new perspective and inspire new directions. More importantly, journal publication involves lengthy review and publication delays. If the ICA annual conference data is public and updated regularly, scholars will find it easier to get the most recent research ideas that are not found in journal publications.

- Circumvent publication biases. Publications might have biases (Sun & Pan, 2020). Not all research projects end up being published. For example, significant results or large effect sizes make studies more likely to be published (Sun & Pan, 2020). Publication bias suggests that we might need to include unpublished works such as conference papers. The papers of the ICA annual conference papers serve as a good starting point. Although these presentations are not publications per se, and therefore not ready to be cited, they are still peer-reviewed, ensuring the quality of the paper presented each year.
- Enable large scientometric analysis. The ICA annual conference data over the past two to three decades are large. It contains 27,466 papers, 21,038 authors, and 4,935 sessions. This dataset is useful for large scale scientometric analysis. For example, the conference-paper dataset can be applied to study the topic evolution of communication studies in the past decades or to study academic collaboration or mobility within the field of Communication.
- Contribute to open science. By making the data publicly available, scholars from all other fields can use this dataset. This dataset also allows communication scholars to build on each other's work more efficiently and build trust in scientific integrity.
- Provide deeper insights. With these data, we can understand the diversity of communication scholars & research topics better. Right now, we only have access to journal data, but that is only part of communication scholars and communication research. To get a broader picture and a deeper understanding, we need data about the conference papers as well.

Motivated by these strengths, benefits, and potential contributions to the whole communication field and broader academia, we obtained and processed all available data on ICA annual conferences. An API for the data and a user interface to help users navigate the data were also developed.

## Related Work

Scientometric and bibliometric research has documented the evolving intellectual landscape of communication scholarship across multiple dimensions (Bryant & Miron, 2004; Chan & Grill, 2022; Walter et al., 2018). The annual number of articles published in high-impact communication journals more than doubled between 2000 and 2017 (Chan & Grill, 2022). U.S. scholars used to contributed 100% of articles in the *Journal of Communication* during the 1950s, while the share declined to 67.4% by the 2010s (Walter et al., 2018), and the analyses of ICA annual conferences indicate that country-level diversity more than doubled between 2005 and 2022 (Scharkow & Trepte, 2024). Yet this internationalization remains structurally uneven such that North American scholars contributed 45.7% of the 123,990 articles published between 1990 and 2019, while Latin America, Southeast Asia, and Sub-Saharan Africa together accounted for less than 8% of all publications (Ekdale et al., 2022). Meanwhile, journals from the Global South face persistent pressure to conform to Northern academic standards as a prerequisite for database indexing (Marques et al., 2026).

Turning to the topical foci of the field, Doerfel & Barnett (1999) conducted a semantic network analysis of 703 paper titles from the 1991 ICA annual conference, finding a significant correlation between the co-word-based semantic network and the division-membership affiliation network, revealing three principal dimensions structuring the association's divisions: humanistic versus scientific orientations, mediated versus interpersonal communication, and theoretical versus applied research. Later work turned to computational topic modeling to examine topical patterns at greater scale. Chan & Grill (2022) conducted a topic modeling analysis of nearly 13,000 abstracts from 18 high-impact journals and identified 40 distinct research clusters. Among these, media effects, social media, persuasion, media psychology, and political communication commanded the highest citation prestige, while social media, clinical communication, media effects, and health campaigns exhibited the most pronounced growth in publication volume over time (Chan & Grill, 2022). Uses and gratifications, cultivation theory, and agenda-setting theory were once the most frequently cited across major journals in mass communication (Bryant & Miron, 2004), but framing theory has since emerged as the leading framework in the *Journal of Communication*, alongside growing attention to social cognitive theory, third-person effects, and social identity theory (Bryant & Miron, 2004; Walter et al., 2018).

Despite these developments, the field's methodological and paradigmatic structure has remained largely unchanged. Post-positivist paradigms accounted for approximately 80% of all articles published in the *Journal of Communication* between 1951 and 2016 (79.8%), with quantitative methods displaying a similarly dominant pattern (79.1%) (Walter et al., 2018). Critical, cultural, and rhetorical approaches peaked in the 1980s and 1990s, reaching approximately 36.2% in the latter decade before declining steadily thereafter (Walter et al., 2018). At the level of citation structure, each research topic tends to be concentrated in a dominant journal (Chan & Grill, 2022), but the top 20% of publications absorb over 70% of all citations (Chan & Grill, 2022), and cross-subdisciplinary co-authorship increased only marginally over sixty-five years, from 20.4% to 25.2% (Walter et al., 2018). These structural asymmetries in the literature are manifest not only at research output and citation distribution, but also in the demographic composition of the scholarly community itself.

At the level of race, Chakravartty et al. (2018) pioneered the `#CommunicationSoWhite` critique, drawing on publication data from 12 communication journals spanning 1990 to 2016 to document the pervasive underrepresentation of non-White scholars in both publication rates and citation counts: non-White authors received an average of 16 citations compared to 25 for White authors. They introduced the concept of "citational segregation," demonstrating that White authors cite other White authors at rates significantly exceeding chance expectation. Building on this, Hatfield et al. (2024) conducted an intersectional analysis of 11,292 authors across five ICA flagship journals over seven decades, finding that White authors constituted approximately 82.4% of all authorships while Black authors accounted for only 1.4%. A demographic analysis of the communication citation elite (CCE) further revealed that among the most citation-influential scholars, 91.5% are White, 74.3% are men, and 78.6% are employed at U.S. institutions, with Black scholars' representation declining rather than improving over the twenty-year study period (Freelon et al., 2023).

At the level of gender, an analysis of 86,719 ICA submissions between 2005 and 2022 found that women's participation at the conference stage has reached near-parity with men (59% in 2022), and that years with available childcare services were associated with an average 2% increase in female participation (Braun et al., 2023). Yet these figures may obscure a systematic leaky pipeline that women's representation declines progressively from 55% of conference authors to 43% among journal authors (Trepte & Loths, 2020), 36% among senior faculty (Bothwell et al., 2022), and ultimately 26% within the citation elite (Freelon et al., 2023).

The extent to which racial issues receive scholarly attention within communication research has also been examined. A keyword-based analysis of nine journals spanning 1991 to 2021 found that political communication and empirical journals devoted significantly less attention to race-related topics than did general-interest and critical outlets (Freelon et al., 2023). An analysis of 39,795 presentations at National Communication Association (NCA) annual conferences between 2005 and 2024 Dixon et al. (2026) further revealed that only 9.37% is race-related content. Furthermore, race-focused presentations within large divisions were concentrated in the opening and closing days of the conference and in early-morning time slots.

Structured datasets and research infrastructure built around academic conferences have proven valuable for such analyses, with examples emerging from other fields (Isenberg et al., 2017; Lange, 2024). Besides, conference papers are particularly well-suited for tracking disciplinary change: unlike journal articles, which involve lengthy review cycles, conference submissions reflect scholarly priorities in near real time. Yet few dedicated datasets or accessible interfaces have been developed to support systematic research on ICA conference papers. The present project aims to aggregate, analyze, and present ICA conference publication data in a form that is both searchable and interactive.

# Data Collection and Processing

The official website of ICA provides information about all past annual conferences at https://www.icahdq.org/page/annual-conference.

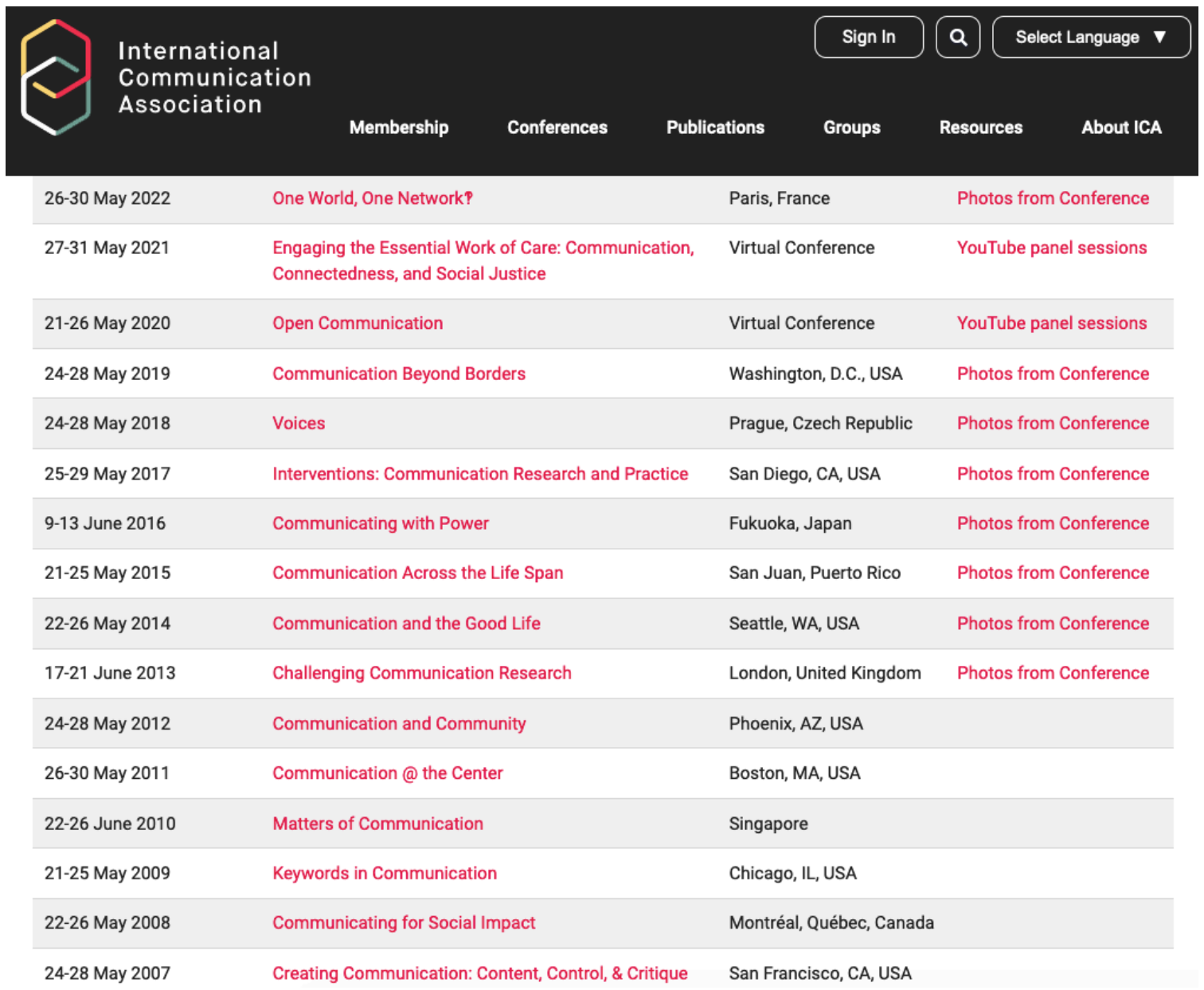

| | | | |
|---|---|---|---|
| 26-30 May 2022 | One World, One Network? | Paris, France | Photos from Conference |
| 27-31 May 2021 | Engaging the Essential Work of Care: Communication, Connectedness, and Social Justice | Virtual Conference | YouTube panel sessions |
| 21-26 May 2020 | Open Communication | Virtual Conference | YouTube panel sessions |
| 24-28 May 2019 | Communication Beyond Borders | Washington, D.C., USA | Photos from Conference |
| 24-28 May 2018 | Voices | Prague, Czech Republic | Photos from Conference |
| 25-29 May 2017 | Interventions: Communication Research and Practice | San Diego, CA, USA | Photos from Conference |
| 9-13 June 2016 | Communicating with Power | Fukuoka, Japan | Photos from Conference |
| 21-25 May 2015 | Communication Across the Life Span | San Juan, Puerto Rico | Photos from Conference |
| 22-26 May 2014 | Communication and the Good Life | Seattle, WA, USA | Photos from Conference |
| 17-21 June 2013 | Challenging Communication Research | London, United Kingdom | Photos from Conference |
| 24-28 May 2012 | Communication and Community | Phoenix, AZ, USA | |
| 26-30 May 2011 | Communication @ the Center | Boston, MA, USA | |
| 22-26 June 2010 | Matters of Communication | Singapore | |
| 21-25 May 2009 | Keywords in Communication | Chicago, IL, USA | |
| 22-26 May 2008 | Communicating for Social Impact | Montréal, Québec, Canada | |
| 24-28 May 2007 | Creating Communication: Content, Control, & Critique | San Francisco, CA, USA | |

Figure 2: Past ICA annual conferences.

Online programs powered by `allacademic.com` exist for the years between 2003 and 2018. Conferences from 2019 onward provide programs in PDF format. These PDFs are very large and hard to parse. Therefore, we decided to leave them alone and focus on online programs instead.

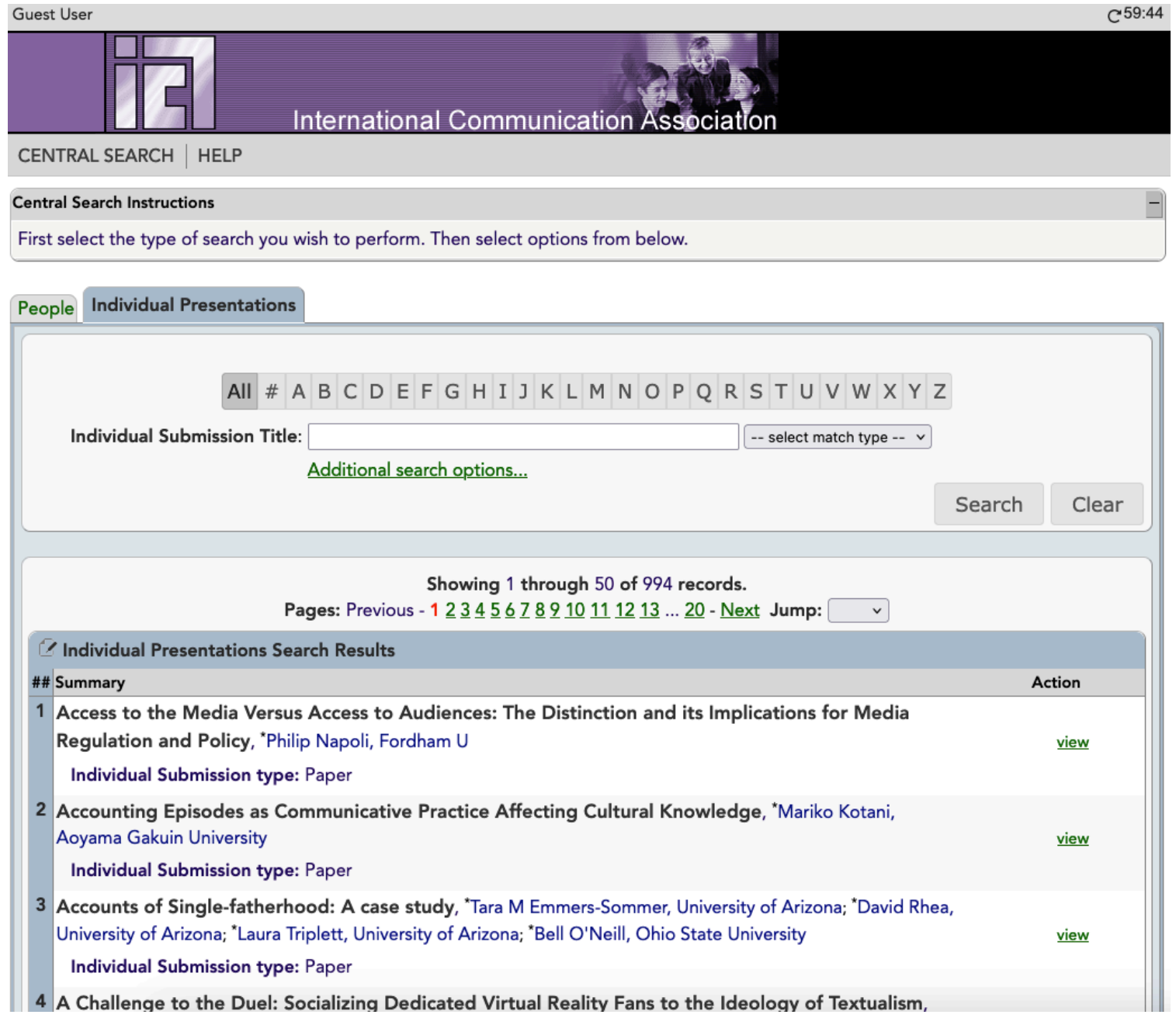

Figure 3: Online program example of ICA annual conference in 2003.

Figure 3 is an example of these online programs. These programs experienced three different periods:

- Years 2003-2004. In these two years, the online programs contain data about authors and papers. However, session and division information is not provided.
- Years 2005-2013. In these years, information about authors, sessions, and papers is provided.
- Years 2014-2018. The online programs changed their structures dramatically, and incorporated Interactive Papers.

We used Selenium to dynamically scrape the website. After data wrangling, three datasets are the main outputs.

The paper data consists of the following columns:

- **Paper ID**: An identifier assigned to each paper, formatted as `year-index`.
- **Title**: The title of the conference paper.
- **Paper Type**: Indicates the type of presentation, either `Paper` or `Poster`. Note that the ICA website did not differentiate between these two types prior to 2014, so all presentations before 2014 are classified as `Paper`, although some may have originally been `Poster`.
- **Abstract**: The abstract of the paper.
- **Number of Authors**: The number of authors for this paper.
- **Year**: The year the paper was presented.
- **Session**: The title of the session in which the paper was presented.
- **Division/Unit**: The division or unit that organized the session.
- **Authors**: The authors of the paper.

The author data includes the following columns:

- **Paper ID**: An identifier assigned to each paper, formatted as `year-index`.
- **Paper Title**: The title of the conference paper.
- **Year**: The year the paper was presented.

- **Number of Authors**: The number of authors for this paper.
- **Author Position**: The position of this author (e.g., first author, co-author).
- **Author Name**: The name of the author.
- **Author Affiliation**: The affiliation of the author.

The session data contains the following columns:

- **Year**: The year the session occurred.
- **Session Type**: Specifies whether the session is a `paper session` or an `interactive paper session` (i.e., poster session).
- **Session Title**: The title of the session.
- **Division/Unit**: The division or unit organizing the session.
- **Chair Name**: The name of the session chair.
- **Chair Affiliation**: The affiliation of the session chair.

These data are available to be downloaded in `CSV` format. There are 27,466 papers authored by 21,038 scholars. These papers are presented in 4,935 sessions.

## Yearly Trends

We analyzed the temporal trends in the number of papers, authors, and average number of authors per paper. Four papers did not have the author information and were excluded from analyses involving authors. Also note that we did not do author deduplication, so authors of the same name are considered the same person even if they are different.

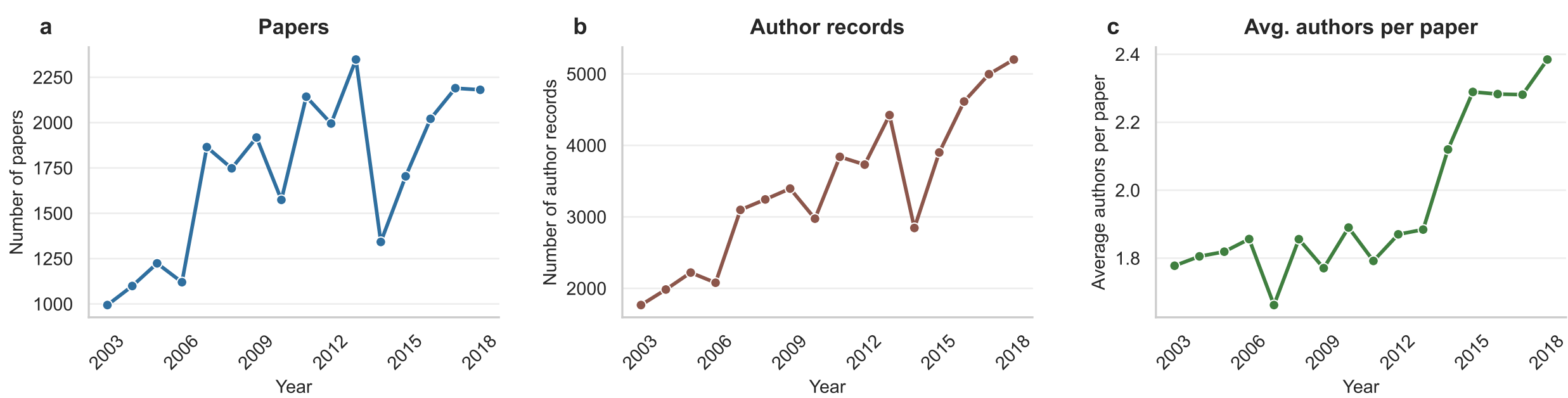

Figure 4: Annual trends in ICA conference papers and authorship (2003–2018). (a) Number of papers by year; (b) number of author records by year; (c) average number of authors per paper by year.

As Figure 4 (a) shows, The number of papers rose from 994 in 2003 to a pre-2014 high of 2,348 in 2013, an increase of about 136%. Growth is uneven, with noticeable declines in 2006, 2008, 2010, and 2012, and a particularly large jump from 2006 to 2007. The sharpest discontinuity is from 2013 to 2014, when the count drops from 2,348 to 1,342 (-42.8%). After that drop, the series recovered to 2,181 papers in 2018, nearly matching 2017 but still below the 2013 peak.

Author records (Figure 4 (b)) increased from 1,767 in 2003 to 4,424 in 2013, a rise of about 150%. Like the paper series, author records declined sharply in 2014, falling to 2,845. The author series then rebounds more strongly than the paper series: by 2018, it reaches 5,201 author records, the highest value in the processed data. This means total authorship activity continues growing even after paper counts level off near 2017-2018.

The average-authors-per-paper trend is the clearest structural change (Figure 4 (c)): papers become more collaborative in the later years, moving from roughly two or fewer authors per paper before 2014 to more than two authors per paper afterward. Because this analysis uses processed files only, the 2013-2014 discontinuity should be interpreted as an observed data break unless confirmed against collection or conference-program changes.

## Topic Modeling

To discover the latent thematic structure of the abstracts of accepted papers from the ICA annual conference between 2003 and 2018 and trace how research interests changed over time, we applied BERTopic (Grootendorst, 2022) to the collection of paper abstracts.

*Text Preprocessing.* BERTopic operates in two sequential stages—document embedding and topic-word representation—each with differing textual requirements: transformer-based embedding models perform optimally on natural, unprocessed text, while topic-word representation benefits from lemmatized text that reduces morphological variants to their base forms. Two versions of the corpus were therefore prepared: the original corpus for embedding and a lemmatized corpus for topic-word representation.

To prepare the lemmatized corpus, abstracts were first cleaned by converting text to lowercase, removing URLs, single-character tokens, numbers, pronouns, and punctuation, and collapsing whitespace. The cleaned texts were then processed using spaCy's `en_core_web_sm` model to reduce each word to its base form (Honnibal et al., 2020). documents that were empty after lemmatization were excluded from both corpora to maintain alignment between the original and lemmatized versions.

*Model Configuration.* Document embeddings were generated from raw abstracts using the `all-mpnet-base-v2` sentence transformer model (Reimers & Gurevych, 2019), and the lemmatized abstracts were used for model training. BERTopic requires parameter selection across three components: dimensionality reduction via UMAP, density-based clustering via HDBSCAN, and token representation via CountVectorizer. We conducted a grid search over 42 parameter combinations (see Table 2 in the appendix) to identify a configuration that produces meaningful topic clusters while minimizing outliers. The final configuration used UMAP with `n_neighbors=20`, `n_components=5`, `min_dist=0.0`, and cosine similarity for dimensionality reduction, and HDBSCAN with `min_cluster_size=200`, `min_samples=5`, and the EOM cluster selection method for clustering. A CountVectorizer with `min_df=10` and unigram-to-bigram range was applied for token representation. This configuration produced 34 topics with an outlier rate of 31.1%.

*Outlier Reduction.* In BERTopic, documents that do not meet the minimum density requirements of the HDBSCAN clustering algorithm are designated as outliers. Following initial model training, outlier reduction was conducted by reassigning each outlier to its nearest topic using cosine similarity, with a threshold of 0.3. This reduced the final outlier rate to 0.2% (54 documents).

*Topic Consolidation.* The 34 model-generated topics were then manually reviewed and consolidated into 15 thematic categories after reassigning one topic consisting predominantly of geographic place names to the outlier category. Top keywords for each topic were then recomputed and filtered using an expanded stopword list including standard English stopwords and domain-specific terms (e.g., study, results, approach). The keywords were subsequently lemmatized, deduplicated, and manually refined by retaining bigrams over their constituent unigrams (e.g., video game over game) and removing morphologically redundant variants. As a result, the 10 most representative keywords for each category are presented in Table 1.

| ID | Topic Category | Representative Keywords |
|---|---|---|
| 1 | Political Communication, Civic Engagement & Activism | political, news, social, public, online, effect, election, information, participation, opinion |
| 2 | Gender, Culture & Identity | woman, cultural, sexual, identity, culture, music, body, gender, television, film |
| 3 | Communication Technology & Digital Media | social, internet, user, technology, mobile, online, digital, network, relationship, facebook |
| 4 | Health Communication | health, message, patient, information, cancer, behavior, effect, risk, intention, attitude |
| 5 | Journalism & News Studies | news, journalism, journalist, newspaper, war, coverage, frame, story, content, press |
| 6 | PR, Crisis & Organizational Communication | organization, public relation, crisis, employee, corporate, work, management, practice, network, company |
| 7 | Interpersonal, Relational & Parasocial Communication | relationship, narrative, character, experience, story, message, perceive, conflict, partner, emotion |
| 8 | Intercultural & Global Communication | china, cultural, intercultural, identity, india, culture, woman, international, language, discourse |
| 9 | Media Policy, Platform Governance & Privacy | privacy, policy, public, service, internet, information, technology, market, network, broadcasting |
| 10 | Video Game & Media Violence | video game, player, play, virtual, presence, effect, experience, violent, behavior, computer |
| 11 | Advertising Effects & Online Consumer Behavior | brand, effect, advertising, message, product, consumer, information, processing, attitude, cognitive |
| 12 | Environment & Science Communication | climate change, environmental, science, frame, issue, risk, news, coverage, food, scientist |
| 13 | Children, Family Media Use & Parental Mediation | child, parent, television, mediation, effect, exposure, violence, parental, view, age |
| 14 | Speech & Rhetoric Communication | language, discourse, interaction, action, speech, repair, speaker, talk, communicative, practice |
| 15 | Sports Media | sport, athlete, fan, team, game, female, event, woman, coverage, gender |

Table 1: BERTopic 15-Topic Taxonomy — ICA Conference Abstracts (2003–2018)

*Temporal Trends.* Figure 5 presents both the overall distribution of the 15 identified topic categories from 2003 to 2018 and their distribution over time. Across the entire period, the six most dominant topics were: (1) political communication, civic engagement, and activism; (2) gender, culture, and identity; (3) communication technology and digital media; (4) health communication; (5) journalism and news studies; and (6) public relations, crisis, and organizational communication. Across the six largest topic categories, Political Communication, Civic Engagement & Activism showed the most significant increase, rising from 9.8% in 2003 to 18.8% in 2018, with a sharp rise after 2015 that likely reflects heightened scholarly attention to the 2016 U.S. presidential election and broader issues such as misinformation and political polarization. In 2014, Journalism & News Studies fell to 7.8% and showed a slight increase in later years, which may also be connected to significant political and technological changes after 2016 that raised research interest in news credibility, misinformation and evolving news environments. Communication Technology & Digital Media increased from 10.1% in 2003 to a peak of 13.0% in 2013, after which it remained relatively stable. This trend likely reflects increasing scholarly attention to digital platforms, mobile communication, and social media. Health Communication fluctuated between 8% and 13% throughout the study period and reached 13.2% by 2018. Public Relations, Crisis, and Organizational Communication fluctuated during the first three years and then remained relatively stable, rising to around 11% in 2014 and holding steady thereafter.However, Gender, Culture & Identity dropped from 15.0% to 9.1% within the same period.

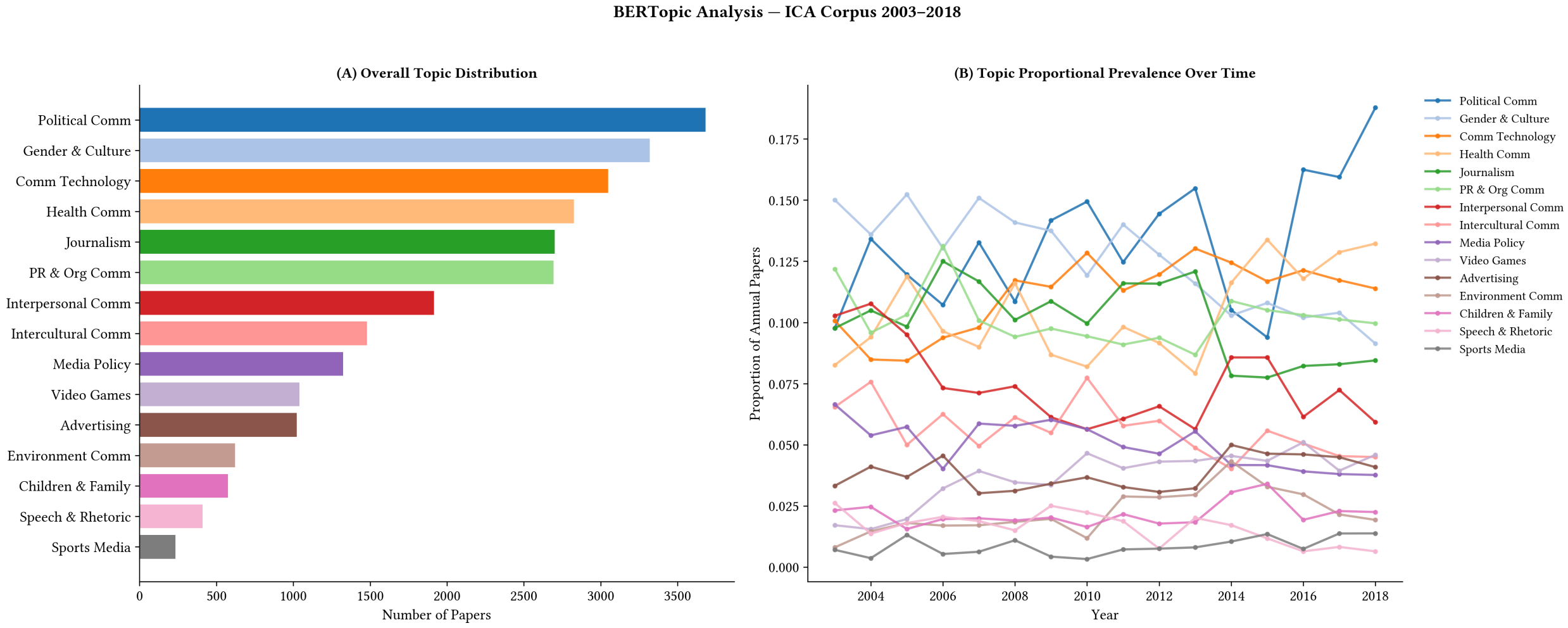

Figure 5: Distribution and temporal trends of 15 BERTopic-identified thematic categories across ICA conference papers (2003–2018). (A) Overall number of papers per topic category; (B) proportional prevalence over time.

## Semantic Search

We added the function of semantic search. The motivation is that keyword search is limited. Sometimes, users will want to find semantically relevant papers to their long input paragraph. For example, “I am interested in how social media misinformation circulates during elections, how people correct false claims, and whether partisan identity shapes what audiences believe.” Keyword search performs poorly for such long natural-language queries.

We created embeddings using Openai’s `text-embedding-3-small` model for all 27,466 papers using their title and abstracts. For the 23 papers without abstracts, we only used their titles. We then stored these vectors on Pinecone, a vector database. When users enter their search query, we use `text-embedding-3-small` to create its embedding and use cosine similarity to retrieve the top relevant papers.

## User Interface

Clearly, the data is very large with around 30,000 papers and 20,000 authors. To make the data more easily available to the public and interested scholars, we developed a web application using `React`.

Figure 1 shows the basic interface of this web app. The left panel shows all the 27,466 papers and the right panel shows the details of each paper.

The interface also includes authors as in Figure 8 and sessions as in Figure 9. The Authors page shows all the 21,038 authors. Clicking on the author will show all the papers by that author. Similarly, the Sessions page shows all the 4,935 sessions and clicking on each session will go to the page that presents all the papers in that session.

In the paper panel, filtering through the year is enabled.

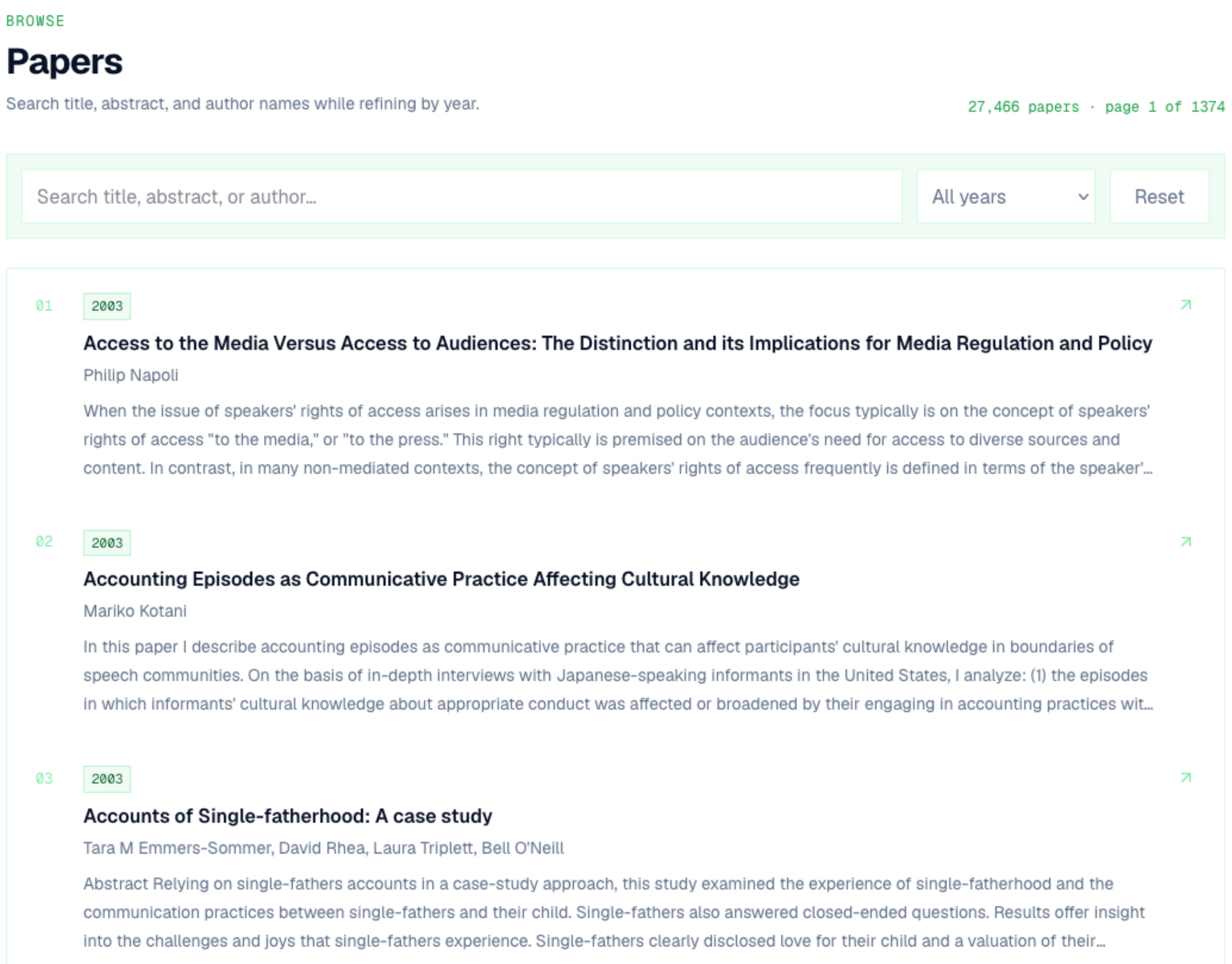

Figure 6: Screenshots of the Papers page in the user interface

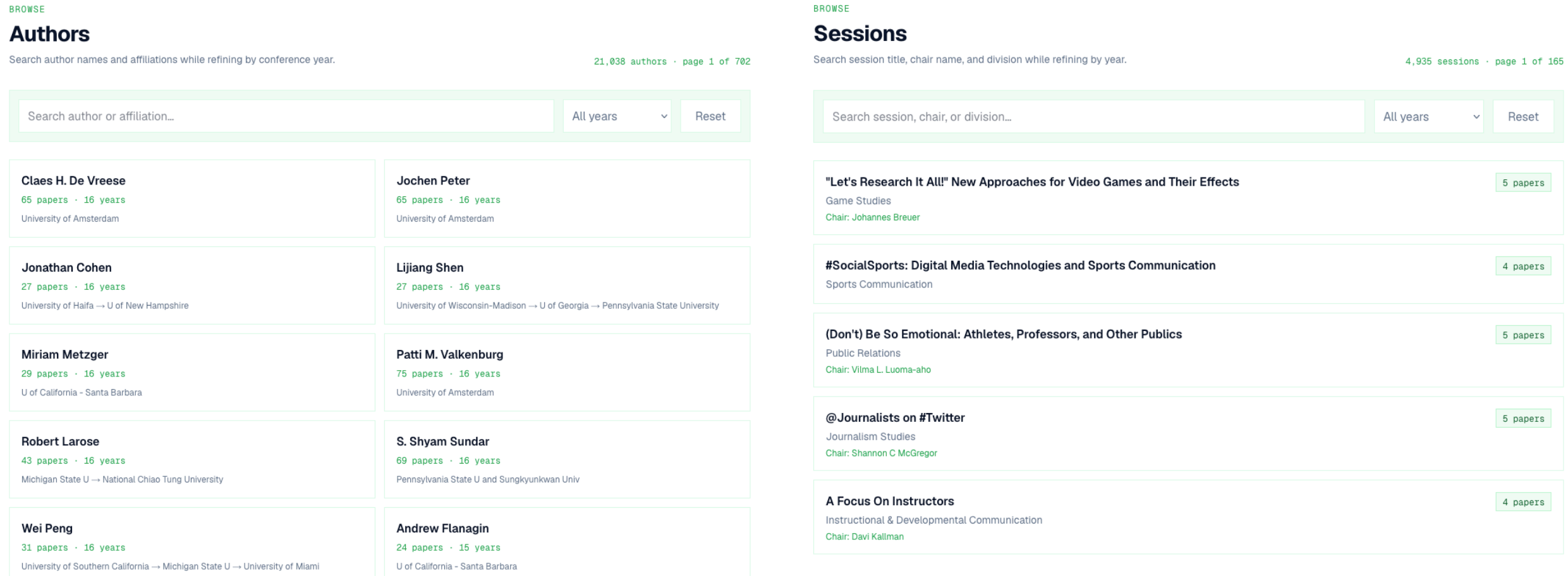

Figure 8: Author Page

Figure 9: Session Page

Figure 7: Screenshots of the Author and Session Pages in the user interface.

## Discussion and Contribution

This paper presents a large and comprehensive dataset of the past ICA annual conference papers from 2003 to 2018, encompassing 27,466 papers, 21,038 authors, and 4,935 sessions. The contributions of this paper are as follows:

- Created a comprehensive dataset that aggregated all past ICA conference papers, authors, and sessions covering 2003 to 2018, which, to our knowledge, is the first work in this direction.
- Developed an API to allow other developers and scholars to utilize the data more easily.
- Designed an interface to allow easier navigation through, and better presentation of, the dataset.

The dataset is open to the public via a live website interface and API. Scholars in the communication field or other fields can easily use this dataset to get deeper insights into how communication research has evolved over the past years, the research topics/focus/subfields evolution trend, get inspiration from past conference papers, or conduct more comprehensive large-scale scientometric analysis.

## Limitations

There are several directions for future improvement. For example:

- Data from years after 2019 and before 2003 are not available yet.
- Deduplication of the author names, distinguishing between two different authors who share the same name, is almost impossible. Usually, we can rely on disciplines to deduplicate authors. However, in this dataset, all authors belong to Communication. Also, many contributors are students who move more frequently between institutions, which makes the deduplication task even harder.
- We did not deduplicate affiliations.

## Contributions

- Hongtao Hao: Initiated the project; scraped, cleaned, organized, and analyzed the data; developed the web interface; did paper embedding and vector store for semantic search; developed the API; wrote and reviewed the paper
- Xinyue Chen: Developed the early version of API in Next.js and semantic search.

- Jiye Sun: Wrote the related work section; Completed the topic modeling and wrote the section in the paper.
- Yanling Zhao: Reviewed and contributed to part of the paper.
- Jing Zhang: Reviewed and contributed to part of the paper.

# Appendix

This appendix details the API development.

## A. Aggregation and Data Structure

The strength of data in `CSV` is that they are lightweight and easy to parse and analyze. However, the drawbacks are obvious. First, it is not friendly for development. In web development, pagination is preferred when the data is very large. However, it is not easily ready with `CSV`. Second, `CSV` data makes it hard to use nested structures. For example, for each paper, it is hard to combine paper data with author data. Suppose I want to add all the author names and author affiliations, it is hard to implement in `CSV`.

Therefore, we decided to aggregate all three datasets, i.e., `papers.csv`, `authors.csv`, and `sessions.csv` into one single data in `JSON` format, which is optimal for nested data structure.

This aggregated data is `papers.json`. Its structure is as follows:

```
class Authorship(BaseModel):
    position: Optional[int] = None
    author_name: Optional[str] = None
    author_affiliation: Optional[str] = None

class SessionInfo(BaseModel):
    session: str
    session_type: Optional[str] = None
    chair_name: Optional[str] = None
    chair_affiliation: Optional[str] = None
    division: Optional[str] = None
    years: List[int] = []
    paper_count: Optional[int] = None
    session_id: Optional[str] = None

class Paper(BaseModel):
    paper_id: str
    title: str
    paper_type: str
    abstract: Optional[str] = None
    number_of_authors: int
    year: int
    session: Optional[str] = None
    division: Optional[str] = None
    authorships: Optional[List[Authorship]] = None
    author_names: Optional[List[str]] = None
    session_info: Optional[SessionInfo] = None
```

As can be seen, it combines both the author and session data nicely.

To facilitate analysis, we also aggregated the author data and session data:

```
class Author(BaseModel):
    author_name: str
    attend_count: int
    paper_count: int
    paper_ids: Optional[List[str]] = None
    affiliations: Optional[List[str]] = None
```

```python
    affiliation_history: Optional[str] = None
    years_attended: Optional[List[int]] = None

class Session(BaseModel):
    session: str
    session_type: Optional[str] = None
    chair_name: Optional[str] = None
    chair_affiliation: Optional[str] = None
    division: Optional[str] = None
    years: Optional[List[int]] = []
    paper_count: Optional[int] = None
    session_id: str
```

The data is served on MongoDB.

## B. API

To make the data easier to use for web development, we developed an API using `Next.js`. Below, we explain the most important endpoints.

### Retrieve Papers

**Endpoint:** `GET /papers`

**Description:** Retrieves a list of papers with optional filters for searching by various fields.

**Parameters:**

- `page` (int): Page number for pagination (default: 1).
- `limit` (int): Number of items per page (default: 100).
- `paper_id` (str): Unique ID assigned to the paper.
- `title_contains` (str): Keyword to search within the paper title.
- `paper_type` (str): Type of presentation, either `Paper` or `Poster`.
- `abstract_contains` (str): Keyword to search within the paper abstract.
- `number_of_authors` (int): Number of authors.
- `session_contains` (str): Keyword to search within the session title.
- `year` (int): The year the paper was presented.
- `session` (str): The session title.
- `division` (str): Division or Unit that organized the session.
- `has_author` (str): Author name appearing in the paper.
- `first_author` (str): First author of the paper.
- `last_author` (str): Last author of the paper.
- `session_id` (str): Exact match for the session ID.

**Example Request:**

```
GET /papers?title_contains=communication&year=2003
```

**Example Response:**

```
[
    {'paper_id': '2003-0155',
     'title': 'Computer-Mediated Communication,...',
     'paper_type': 'Paper',
     'abstract': 'The present study ...',
```

```
     'number_of_authors': 2,
     'year': 2003,
     'session': None,
     'division': None,
     'authorships': [{'position': 0,
                      'author_name': 'Jo Anna Madrid',
                      'author_affiliation': 'Rio Hondo College'},
                     {'position': 1,
                      'author_name': 'Richard L. Wiseman',
                      'author_affiliation': 'California ...'}],
     'author_names': ['Jo Anna Madrid', ...],
     'session_info': None},
    ...
]
```

**Retrieve a Paper by ID**

**Endpoint:** `GET /papers/{paper_id}`

**Description:** Retrieves detailed information for a specific paper by its unique `paper_id`.

**Parameters:**

- `paper_id` (str): The unique ID of the paper.

**Example Request:**

```
GET /papers/2003-001
```

**Example Response:**

```
[{'paper_id': '2003-0001',
  'title': 'Access to ...',
  'paper_type': 'Paper',
  'abstract': 'When the issue ...',
  'number_of_authors': 1,
  'year': 2003,
  'session': None,
  'division': None,
  'authorships': [{'position': 0,
                   'author_name': 'Philip Napoli',
                   'author_affiliation': 'Fordham U'}],
  'author_names': ['Philip Napoli'],
  'session_info': None}]
```

**Retrieve Authors**

**Endpoint:** `GET /authors`

**Description:** Retrieves a list of authors with optional filters to search by various fields.

**Parameters:**

- `page` (int): Page number for pagination (default: 1).
- `limit` (int): Number of items per page (default: 50).
- `author_name` (str): Exact name of the author.
- `min_attend_count` (int): Minimum number of times the author attended.
- `min_paper_count` (int): Minimum number of papers the author has.

- `affiliation_contains` (str): Keyword to search within author affiliations.
- `year_attended` (int): Specific year the author attended.

**Example Request:**

```
GET /authors?author_name=Eiri Elvestad
```

**Example Response:**

```
[{'author_name': 'Eiri Elvestad',
  'attend_count': 2,
  'paper_count': 2,
  'paper_ids': ['2013-1099', '2018-0124'],
  'affiliations': ['Vestfold U College', 'U ...'],
  'affiliation_history': 'Vestfold U College -> U ...',
  'years_attended': [2013, 2018]}]
```

**Retrieve Sessions**

**Endpoint:** `GET /sessions`

**Description:** Retrieves a list of sessions with optional filters for specific session attributes.

**Parameters:**
- `page` (int): Page number for pagination (default: 1).
- `limit` (int): Number of items per page (default: 50).
- `session` (str): Exact name of the session.
- `session_type` (str): Type of the session, e.g., `Paper Session`.
- `chair_name` (str): Name of the session chair.
- `chair_affiliation` (str): Affiliation of the session chair.
- `division` (str): Division or unit organizing the session.
- `year` (int): Specific year the session was held.
- `paper_count` (int): Number of papers in the session.

**Example Request:**

```
GET /sessions?paper_count=10
```

**Example Response:**

```
[
    {'session': 'Best Student Papers in Public Relations',
     'session_type': 'Paper Session',
     'chair_name': 'Chiara Valentini',
     'chair_affiliation': 'Aarhus U',
     'division': 'Public Relations',
     'years': [2014, 2015],
     'paper_count': 10,
     'session_id': '005831a276e8'
    },
    ...
]
```

This can be easily done in Python. For example:

```python
import requests

base_url = ""

# Parameters for the request
params = {
    "title_contains": "communication",
    "year": 2003
}

# Make the request
response = requests.get(f"{base_url}/papers", params=params)

# Check the response
if response.status_code == 200:
    papers = response.json()
    print("Papers retrieved:", papers)
else:
    print("Failed to retrieve papers:", response.status_code, response.text)
```

## C. BERTopic Parameter Grid Search Results

| n_neighbors | min_cluster_size | min_samples | N Topics | Outlier Rate |
|---|---|---|---|---|
| 15 | 50 | 5 | 128 | 31.6% |
| 15 | 50 | 10 | 118 | 34.7% |
| 15 | 80 | 5 | 81 | 33.3% |
| 15 | 80 | 10 | 76 | 34.6% |
| 15 | 100 | 5 | 64 | 35.0% |
| 15 | 100 | 10 | 2 | 0.2% |
| 15 | 120 | 5 | 55 | 35.0% |
| 15 | 120 | 10 | 2 | 0.2% |
| 15 | 150 | 5 | 44 | 33.7% |
| 15 | 150 | 10 | 2 | 0.2% |
| 15 | 180 | 5 | 39 | 34.1% |
| 15 | 180 | 10 | 2 | 0.2% |
| 15 | 200 | 5 | 2 | 0.2% |
| 15 | 200 | 10 | 2 | 0.2% |
| 20 | 50 | 5 | 118 | 35.2% |
| 20 | 50 | 10 | 112 | 37.0% |
| 20 | 80 | 5 | 81 | 33.8% |
| 20 | 80 | 10 | 74 | 34.5% |
| 20 | 100 | 5 | 59 | 30.5% |
| 20 | 100 | 10 | 65 | 35.3% |
| 20 | 120 | 5 | 53 | 29.1% |
| 20 | 120 | 10 | 58 | 34.8% |
| 20 | 150 | 5 | 45 | 31.7% |
| 20 | 150 | 10 | 2 | 0.2% |
| 20 | 180 | 5 | 37 | 31.6% |
| 20 | 180 | 10 | 2 | 0.2% |
| 20 | 200 | 5 | 34 | 31.1% |
| 20 | 200 | 10 | 2 | 0.2% |
| 25 | 50 | 5 | 114 | 37.0% |
| 25 | 50 | 10 | 103 | 34.2% |
| 25 | 80 | 5 | 72 | 33.9% |
| 25 | 80 | 10 | 69 | 34.4% |
| 25 | 100 | 5 | 61 | 34.9% |

| n_neighbors | min_cluster_size | min_samples | N Topics | Outlier Rate |
|---|---|---|---|---|
| 25 | 100 | 10 | 55 | 33.9% |
| 25 | 120 | 5 | 53 | 35.7% |
| 25 | 120 | 10 | 48 | 34.5% |
| 25 | 150 | 5 | 45 | 34.2% |
| 25 | 150 | 10 | 2 | 0.2% |
| 25 | 180 | 5 | 35 | 34.1% |
| 25 | 180 | 10 | 2 | 0.2% |
| 25 | 200 | 5 | 2 | 0.2% |
| 25 | 200 | 10 | 2 | 0.2% |

Table 2: Parameter combinations evaluated in the BERTopic grid search. All combinations used `min_dist = 0.0`.